\documentclass[aps,prx,twocolumn,longbibliography]{revtex4-1}%

\usepackage{graphicx}
\usepackage{float}
\usepackage{braket}
\usepackage{pslatex,latexsym,times}
\usepackage{bm, float, lipsum}
\usepackage{amssymb}
\usepackage{color}
\usepackage{rotating}
\usepackage[normalem]{ulem}

\begin{document}

\title{Immense magnetic response of exciplex light emission due to correlated spin-charge dynamics}

\author{Yifei Wang}
\thanks{These authors contributed equally to this work}
\author{Kevser Sahin-Tiras$^*$, Nicholas J. Harmon, Markus Wohlgenannt, Michael E. Flatt\'e}\email{michael\_flatte@mailaps.org}
\affiliation{Department of Physics and Astronomy and Optical Science and Technology Center, University of Iowa, Iowa City, Iowa 52242, USA}


\begin{abstract}
 As carriers slowly move through a disordered energy landscape
 in organic semiconductors, tiny spatial variations in spin dynamics relieve spin blocking at transport bottlenecks
 or in the electron-hole recombination process   that produces light.  Large room-temperature magnetic-field effects (MFE) ensue in the conductivity and luminescence.
Sources of variable spin dynamics  generate much larger MFE if their spatial structure is correlated on the nanoscale with the energetic sites governing conductivity or luminescence such as in co-evaporated organic blends within which the electron resides on one molecule and the hole on the other (an exciplex). Here we show that exciplex recombination in  blends exhibiting thermally-activated delayed fluorescence (TADF) produces MFE in excess of 60\% at room temperature. In addition, effects greater than 4000\% can be achieved by tuning the device's current-voltage response curve by device conditioning. These immense MFEs are both the largest reported values for their device type at room temperature. 
 Our theory traces this MFE and its unusual temperature dependence to changes in spin mixing between triplet exciplexes and light-emitting singlet exciplexes. In contrast, spin mixing of excitons is energetically suppressed, and thus spin mixing produces comparatively weaker MFE in materials emitting light from excitons by affecting the precursor pairs. 
Demonstration of immense MFE in common organic blends provides a flexible and inexpensive pathway towards magnetic functionality and field sensitivity in current organic devices without patterning the constituent materials on the nanoscale.  Magnetic fields increase the power efficiency of unconditioned devices by 30\% at room temperature, also showing that magnetic fields may increase the efficiency of the TADF process.
\end{abstract}

\maketitle

\section{Introduction}

Organic light-emitting diodes (OLEDs)\cite{Friend:1999:Nature} based on organic semiconductors are used extensively in flat-panel displays and other lighting, due to their mechanical flexibility and low-cost fabrication. In most OLEDs
electrons and holes are injected into the device and, upon encountering one another, form a loosely bound state (a polaron-pair) and finally a tightly-bound exciton. Because both the electron and hole carry spin-1/2, these bound states can be spin singlets (total spin 0), or triplets (total spin 1), and they usually form in a 1:3 ratio based on spin degeneracy. Most organic semiconductors are fluorescent materials with an internal electroluminescence quantum efficiency $\eta$ limited to $\eta \approx$ 25\%  because  only singlet excitons lead to significant electroluminescence\cite{Rothberg1996}. Recently, triplet-to-singlet up-conversion in thermally-activated delayed fluorescence (TADF) blends has  increased $\eta$ well above 25\% \cite{Goushi2012,Uoyama2012}. TADF requires the exchange splitting between the singlet and triplet states, $\Delta_{ST}$, to be smaller than or comparable to the thermal energy ($\approx$ 25 meV at room temperature).
As $\Delta_{ST}$ in excitons is usually orders of magnitude larger, intermolecular excitations or so-called exciplexes,
which have much smaller $\Delta_{ST}$, are a common choice to achieve TADF.
In parallel to these improvements in OLED emission efficiency, recent investigations have identified large magnetic-field sensitive spin effects on the electroluminescence in organic semiconductors \cite{Kalinowski2003,Mermer:2005:PRB,Prigodin:2006:SynthMet,Desai:2007:PRBphotocurrent,Hu:2007:NatMat,Wang:2012:PRX,Macia2014,Crooker2014,Zhang:2015:NatPhys}. Organic semiconductors also possess an intrinsic magnetoresistive effect, termed organic magnetoresistance (OMAR) which in  magnetic fields, $B$, of just a few mT, is
$\approx$ 20\% at room temperature in a large variety of organic semiconductors\cite{Mermer:2005:PRB,Prigodin:2006:SynthMet}, and is often accompanied by magnetoelectroluminescence (MEL)\cite{Kalinowski2003,Mermer:2005:PRB}.
A small $\Delta_{ST}$ suggests that the spin dynamics of recombination in TADF blends could be dramatically modified by a small magnetic field, producing a very large OMAR effect and improving the efficiency of OLEDs.

 Here we  show an immense ($>50$\%) magnetic field effect occurs to both the resistance and the electroluminescence in an organic  blend  already known to possess a large internal quantum efficiency due to TADF. The dramatic enhancement of this magnetic field effect is due to the properties of the singlet-triplet exchange splitting, $\Delta_{ST}$, which also explains the temperature dependence of the magnetic field effects. The devices were fabricated using typical micron-scale fabrication processes, and do not require any nanoscale structuring of the constituent molecules. Thus this approach appears broadly applicable to a wide variety of organic devices which incorporate TADF blends without complex fabrication. We  experimentally demonstrate in our devices a 30\% increase in the power efficiency of light emission at room temperature due to the application of a magnetic field, compared to the efficiency of the same device at zero field.

A theory that includes the effects of the singlet-triplet level structure associated with TADF is developed and the immense effects due to interaction of these exciplexes with hyperfine fields and $g$ factor variations  are calculated.  The effects of hyperfine fields and $g$ factor variations oppose each other, so by the demonstrated increased emission with magnetic field we identify $g$ factor variations ({\it i.e.}, the $\Delta g$ mechanism \cite{Wang:2008:PRL, DevirWolfman2014}) as the dominant mechanism of the effect. 

 We further demonstrate the ability to ``condition'' the organic devices to the point where the change in luminescence or resistance exceeds an order of magnitude, with the largest changes in excess of 4000\%. Conditioning leads to a decrease in the $\Delta_{ST}$, which enhances the MFE at constant current, and also an undesirable increase in the device resistance. We thus clarify the connection between magnetic-field effects in current and those in voltage, demonstrating that those effects seen in constant voltage can be amplified through various conditioning procedures, whereas those in constant current are less responsive. The size of these magnetic field effects, $>50$\%\ even in unconditioned devices, are the largest reported to date at room temperature.

\section{Theory}

\begin{figure*}
\includegraphics[width=\linewidth]{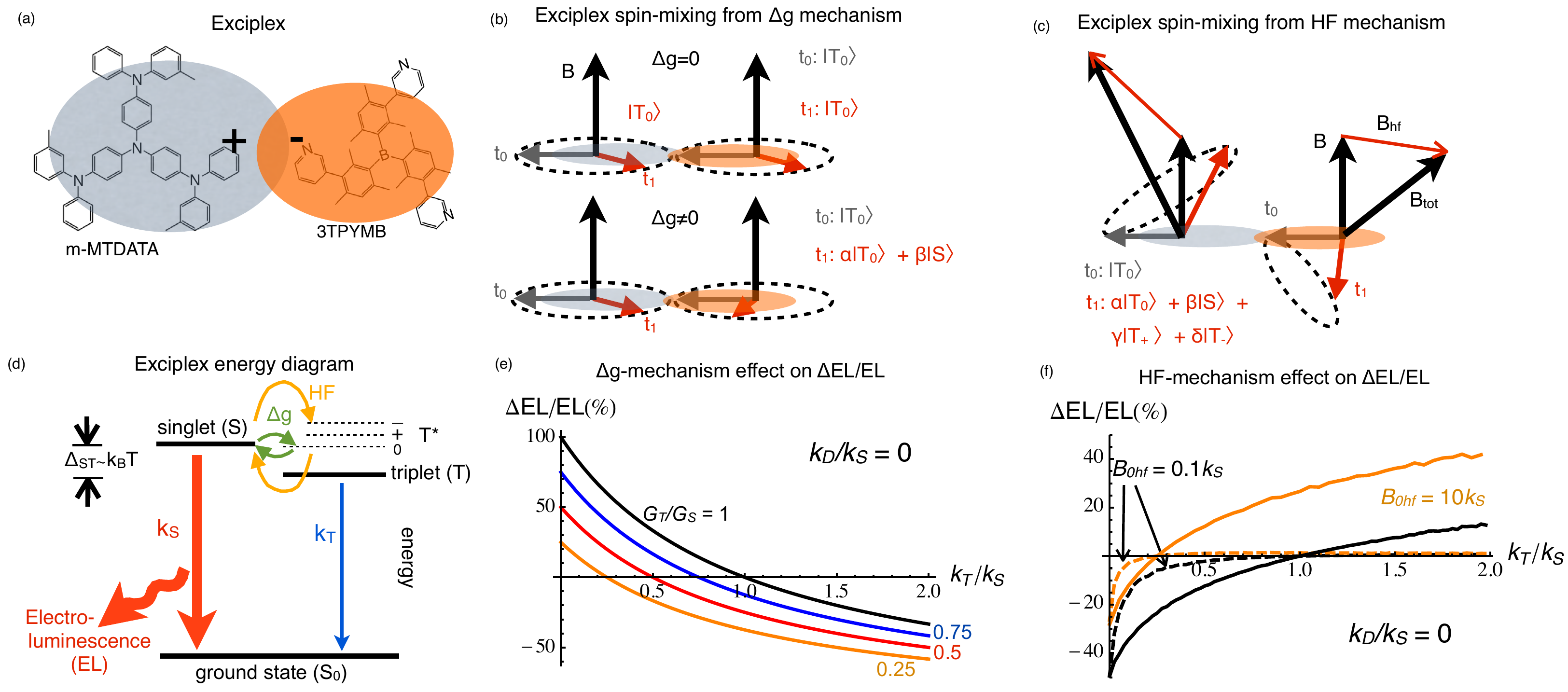}
\caption{\label{fig1}
\sloppy
{Exciplexes,  $\Delta g$ mechanism and hyperfine mechanism for magnetic field effects. \rm{(a) Schematic of the exciplex with m-MTDATA as the donor and 3TPYMB as the acceptor. The electron (hole) wave function is schematically represented by the orange (grey) oval and are overlaid upon the respective chemical structures. (b) $\Delta g$ spin mixing between the electron and hole spins (shown to be orthogonal to the applied field $\bm{B}$) which are initiated at time $t_0$ in the $|T_0\rangle$ spin configuration. In the top panel, where there is no difference in $g$-factors, the spins remain in their initial configuration. In the lower panel, with non-zero $\Delta g$, the spinor picks up singlet ($|S\rangle$) character over time $t_1$. (c) As in (b), but hyperfine spin mixing between electron and hole spins. In this case, spin mixing occurs between all four spin states. (d) Energy diagram of exciplex showing the exchange splitting, $\Delta_{ST}$, between the singlet and triplet levels. For efficient TADF, $\Delta_{ST}$ should be similar in magnitude to the thermal energy $k_B T$. The much quicker singlet recombination (with rate $k_S$) is depicted by the thicker arrow compared to triplet recombination (with rate $k_T$) and luminescence is assumed only for the singlet recombination. Spin-mixing occurs due to the $\Delta g$ mechanism (i.e. transitions between $S$ and the activated $T_0$ level of $T^{\ast}$) and the hyperfine mechanism (i.e. transitions between $S$ and all activated triplet levels of $T^{\ast}$). (e) Theoretical calculations for the MEL$=\Delta EL/EL$ at large field with the $\Delta g$-mechanism ($B_{0hf} = 0$).
(f) Theoretical calculations for the MEL$=\Delta EL/EL$ from hyperfine spin mixing at large field ($\Delta g = 0$). Increasing either $G_T/G_S$ or $B_{0hf}$ causes the MEL to be more negative for exciplex recombination values. Orange lines are $G_T/G_S = 0.25$ and black lines are $G_T/G_S = 1$ . For (e) and (f), $k_D$ is the exciplex dissociation rate, and $G_S$ is the singlet exciplex generation rate, whereas $G_T$ is the generation rate for thermally activated triplet exciplexes (see text for additional details).
}}
}
\end{figure*}

In this section the effects of magnetic fields on spatially-separated entities, such as the electrons and holes composing exciplexes, are contrasted with the effects of magnetic fields on excitons in organic semiconductors, and a theory is developed to describe exciplex magnetic-field effects.
Magnetic field effects  in organic semiconductors occur when spin-carrying (paramagnetic) entities, such as electrons, holes, or triplet excitons react with each other in spin-selective ways to form charge-neutral excitons\cite{Kalinowski2003}, doubly-charged bipolarons \cite{Bobbert:2007:PRL}, trap-coupled carriers\cite{Harmon2014}, or ``trions" \cite{Janssen:2013:NC}. The spin-selective recombination can be influenced by local or global fields, or any other process that produces a magnetic-field-dependent reaction rate. When the reaction involves  exciplexes then  the electrons and holes are located on two distinct molecules, as shown in Fig.~\ref{fig1}(a). In order to enhance the possibilities of exciplex formation these two types of molecules are commonly evaporated together (co-evaporated). For exciplexes the behavior of the electron and hole  spins can differ considerably, as these electronic excitations reside on different molecules.

Spatially inhomogenous (quasi)magnetic fields also lead to spin-mixing\cite{Wang:2008:PRL,Wang:2012:PRX,Macia2014, DevirWolfman2014}, including {\it the $\Delta$B mechanism}, which arises in situations where there a large magnetic field gradients present, leading to a locally varying spin-precession frequency\cite{Wang:2012:PRX,Macia2014}, and {\it the $\Delta$g mechanism} [see Fig.~\ref{fig1}(b)], which is very similar but the variation of the precession frequency is due to different proportionalities of the spin precession to the magnetic field (Land\'e $g$ factor) on neighboring molecules as illustrated in Fig.~\ref{fig1}(b). The $g$ factor  differs from the free electron value because of material-dependent interactions including spin-orbit coupling. The $\Delta$g mechanism has recently been identified as the dominant MFE in several organic and hybrid perovskite systems.\cite{Wang:2008:PRL,Zhang:2015:NatPhys,DevirWolfman2014}

Another major source of spin dynamics is {\it the hyperfine mechanism} [see Fig.~\ref{fig1}(c)]. Every molecule, denoted by $i$, possesses a local hyperfine field, $\bm{B}_{hf,i}$, on the order of mT, due mainly to the magnetic moment of hydrogen nuclei about which the paramagnetic spin will precess. The hyperfine field direction varies randomly from site to site and therefore causes spin-mixing in paramagnetic pairs on neighboring molecules. A spatially uniform externally applied field exceeding the hyperfine strength suppresses this spin mixing and therefore changes the reaction rate between the entities, which can have an amplified effect on transport in the percolative regime\cite{Harmon:2012:PRL}.

TADF exciplex devices are a promising material system for constructing organic semiconductor devices with immense magnetic field effects. In essence, a limit on the magnetic field effect achievable via the so-called exciton mechanism is eliminated in TADF exciplex materials. In the exciton mechanism\cite{Kalinowski2003,Prigodin:2006:SynthMet}, electrons and holes located on neighboring molecules form a polaron pair, and spin mixing through the mechanisms above is permitted. The next step in the evolution of polaron pairs is the spin-conserving formation of an exciton (in which both carriers inhabit the same molecule).  Once the excitons form, because of the large $\Delta_{ST}$, further spin mixing is not energetically allowed. Although the rates for singlet and triplet exciton formation rates are different, i.e. $k_S \neq k_T$,  both channels are spin-allowed exothermic transitions. In contrast,  for TADF exciplex materials in which $\Delta_{ST}$ is of the order of the thermal energy, as shown in Fig.~\ref{fig1}(d),  singlet-triplet transitions readily occur upon thermal activation and exciplexes play the role of polaron pairs described in the exciton mechanism above. In these materials singlet exciplexes can decay radiatively directly to the singlet ground state, whereas triplet exciplexes cannot unless a spin-flip occurs (i.e. phosphorescence). Therefore  the situation is that the singlet channel is spin allowed, whereas the triplet channel is spin ``forbidden". Thus, $k_S \gg k_T$, and a very large MFE should be possible, for both the $\Delta g$ and hyperfine mechanisms.


The theoretical effects of $\Delta g$ ($\equiv g_1 - g_2$) and hyperfine interactions on the spin dynamics and resulting MEL and MC for TADF blends can be calculated from the stochastic Liouville equation for the two-spin density matrix, $\rho$:\cite{Haberkorn1976}
\begin{eqnarray}\label{eq:SLE}
\frac{\partial \rho}{\partial t} &=&  - \frac{i}{\hbar} [H_0 + H_{\Delta g} + H_{hf} + H_{hf, \Delta g},  \rho] {}\nonumber\\
&&{} -\frac{1}{2} \{  k_S P_S + k_T P_T, \rho \} - k_D \rho + G,
\end{eqnarray}
where the Hamiltonians are
\begin{eqnarray}
&&H_0 = \frac{g_1+g_2}{2} \mu_B B \hat{z} \cdot (\bm{S}_1 + \bm{S}_2), \ \ \ H_{\Delta g} = \frac{\Delta g}{2} \mu_B B \hat{z} \cdot (\bm{S}_1 - \bm{S}_2 ),\nonumber\\
 && {}H_{hf} = \frac{g_1+g_2}{2} \mu_B (\bm{B}_{hf,1}\cdot \bm{S}_1 + \bm{B}_{hf,2} \cdot \bm{S}_2 ),\nonumber\\
 && {}H_{hf, \Delta g} = \frac{\Delta g}{2} \mu_B (\bm{B}_{hf,1}\cdot \bm{S}_1 - \bm{B}_{hf,2} \cdot \bm{S}_2 ),
\end{eqnarray}
$P_S$ and $P_T = P_{T,0} + P_{T,+} + P_{T,-}$ are the singlet and (total) triplet projection operators, and the curly braces denote the anticommutator. The two hyperfine fields are drawn from a Gaussian distribution with width $B_{0hf}$. $G$ is a diagonal matrix with elements $G_S$ and $G_{T,+}=G_{T,0}=G_{T,-}=G_T$ for the generation of singlet and triplet exciplexes, respectively. We assume that all three triplet spin eigenstates have the same generation rate. We also assume that the rate of exciplex formation is unchanged by the presence of the magnetic field, and thus $G_S$ and $G_T$ are constant. Dissociation of the exciplex is also included in the theory with a rate $k_D$.
We consider a steady-state condition and solve for the steady state density matrix. EL is estimated from the singlet fraction of exciplexes as EL $= k_S X_S = k_S Tr[P_S \rho]$.

In general, solutions to Equation (\ref{eq:SLE}) must proceed numerically. However, in the limit of negligible hyperfine fields, solutions for the $\Delta g$ induced MEL can be expressed analytically.
In this case we find
\begin{widetext}
\begin{equation}\label{eq:MELeq}
MEL = \left[\frac{G_T (k_D + k_S) - G_S (k_D+k_T)}{G_S(2 k_D + k_S + k_T)}\right] \frac{(\Delta g \mu_B B/\hbar)^2}{{(k_D + k_T) (k_D + k_S)}+(\Delta g \mu_B B/\hbar)^2}.
\end{equation}
\end{widetext}
The values of the MEL at large field for the $\Delta g$ mechanism, from Eq.~(\ref{eq:MELeq}), are shown  in Fig.~\ref{fig1}(e) for $k_D = 0$ and for various values of $G_T/G_S$ and $k_T/k_S$. The situation for exciplex recombination corresponds to $k_T/k_S\ll 1$, which will allow magnetic field effects as large as 100\% if $G_T/G_S\approx 1$, and produces a {\it positive} MEL. For exciton recombination $k_T/k_S\sim 1$, which produces a much smaller magnetic field effect, as seen in Fig.~\ref{fig1}(e). The sign of the MEL for the $\Delta g$ mechanism differs considerably from MEL caused by the hyperfine mechanism, for which a {\it negative} MEL is expected\cite{Kersten2011}. The values of the MEL at large field for the hyperfine interaction are shown in Fig.~\ref{fig1}(f) for $G_T/G_S=1$ and $G_T/G_S= 0.25$, and two values of the hyperfine field. A decrease in the $\Delta_{ST}$ will produce a larger magnitude MEL in the experimentally-relevant regime of $k_T/k_S\ll 1$, whether the mechanism is the $\Delta g$ mechanism or the hyperfine mechanism.

As the exciplex formation rate is directly related to the current through a device, the assumption of a constant formation rate corresponds to assuming a constant-current experimental condition for evaluating the effects of a magnetic field. Under conditions where the resistance changes substantially in the presence of a magnetic field this assumption must be examined anew, and thus the MEL measured under constant voltage conditions (MEL$|_V$), for which the current and resistance changes substantially, will differ from the MEL measured under constant current conditions (MEL$|_I$).

Our theory directly yields a magnetoluminescence response from the $\Delta g$ and hyperfine mechanisms, but, at first sight, does not address the magnetoconductance (MC). As shall be presented below, however, a sizeable magnetoconductance that is smaller but roughly comparable in magnitude (up to a factor of 2-3) is always measured alongside the MEL response. It was  recognized in the early days of OMAR research that the exciton mechanism for MEL (which in this aspect behaves similarly to our exciplex mechanism for MEL) immediately produces a concomitant MC response\cite{Bergeson2008}. In short, the electron and hole densities are large close to their respective injecting electrode. If the electron-hole recombination is efficient (and {\it mutatis mutandis} for inefficient recombination) there will be little spatial overlap of the two densities in the center of the device. This is because electrons and hole will immediately recombine in any part of the device where the two densities overlap. Therefore the majority of the device is electrically charged, limiting the conductance by the so-called space-charge limited current law. Any magnetic-field dependent recombination rate (as in our exciplex mechanism) will change the width of the portion of the device where the two densities overlap and space-charge cancellation occurs, thus producing a change in the conductance. Thus we assume the same functional dependence on magnetic field for the MC as for the MEL for this exciplex theory.

\section{Device Composition and Fabrication} \label{sec:devicefabrication}

\begin{figure}
\includegraphics[width=\linewidth]{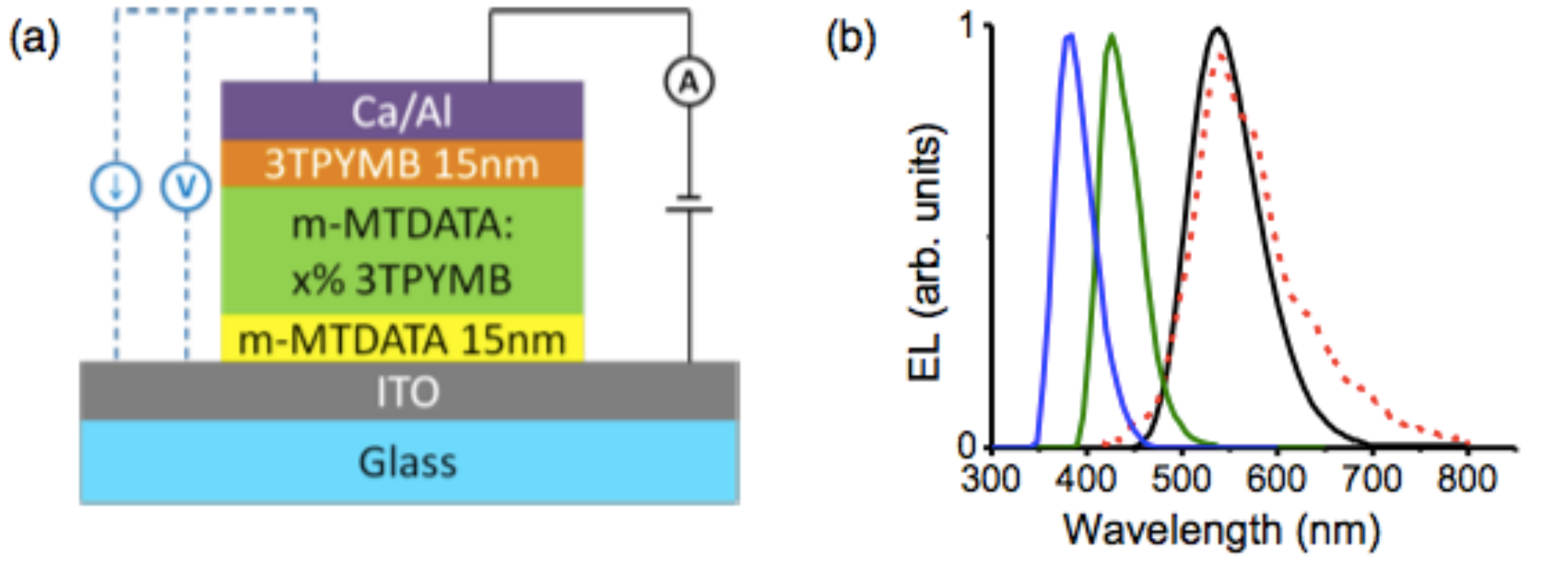}
\caption{\label{fig2ab}
\sloppy
\noindent
 {Exciplex device structure and emission spectrum.}
({a}) Schematic device structure showing the two measurement modes used. The solid circuit is for a constant voltage measurement, whereas the dashed circuit is for a constant current.
({b}) Electroluminescence spectrum for our device (red dashed line), compared with exciplex emission (solid black line) and exciton emission spectra (solid blue and green lines) taken from the literature.\cite{Goushi2012} This shows that exciplexes are responsible for EL in our devices.
}
\end{figure}

Our TADF devices, schematically shown in Fig.~\ref{fig2ab}(a), are thin film devices with several layers deposited sequentially onto a glass slide with patterned indium-tin-oxide (ITO) electrodes. We use a materials combination for the primary layer that is known to produce a large  internal electroluminescence quantum efficiency $\eta$ due to TADF\cite{Goushi2012}:  4,4,4-tris[3-
methylphenyl(phenyl)amino]triphenylamine (m-MTDATA) as donors and tris-[3-(3-pyridyl)mesityl]borane (3TPYMB) as acceptors. 
The m-MTDATA  ($>99\%$ pure) was purchased from Sigma-Aldrich and two batches with different purities of 3TPYMB ($>99.2\%$ pure, used for data in Figs.~\ref{fig6}-\ref{fig4b} and $>99.8\%$ pure, used in all other figures) were purchased from Lumtec (Luminescence Technology Corp.). The materials were used as received.

 After careful cleaning of the ITO/glass substrate in solvents, followed by plasma treatment, the organic layers were fabricated by thermal evaporation under high vacuum  at $10^{-7}$ mbar.
In our devices {the} primary layer consists of co-evaporated m-MTDATA:3TPYMB with a mass fraction of x\% 3TPYMB. All our figures are for x=75\% unless otherwise stated, as this resulted in the largest magnetoelectroluminescence.
 A cathode layer consisting of calcium (covered by a protective layer of aluminum to increase air-stability) was deposited by thermal evaporation in most devices, but e-beam evaporation was used for some devices. The active device area was about $0.6\times0.6$ mm$^{2}$.

 The structure of the devices that gave the largest effects was ITO/m-MTDATA (15 nm)/25 wt\%-m-MTDATA: 3TPYMB (180nm)/3TPYMB (15 nm)/Ca (30 nm)/Al (60 nm). During the co-evaporation of the 25 wt\%-m-MTDATA:3TPYMB, m-MTDATA was deposited at a rate of 0.1~nm/s and 3TPYMB at 0.3~nm/s.
Fig.~\ref{fig2ab}(b) shows the spectrum of the electroluminescence from these layers, compared with emission from excitons in either of the two constituents, and confirms that exciplexes indeed form in our devices\cite{Goushi2012}.

\begin{figure}
\includegraphics[width=\linewidth]{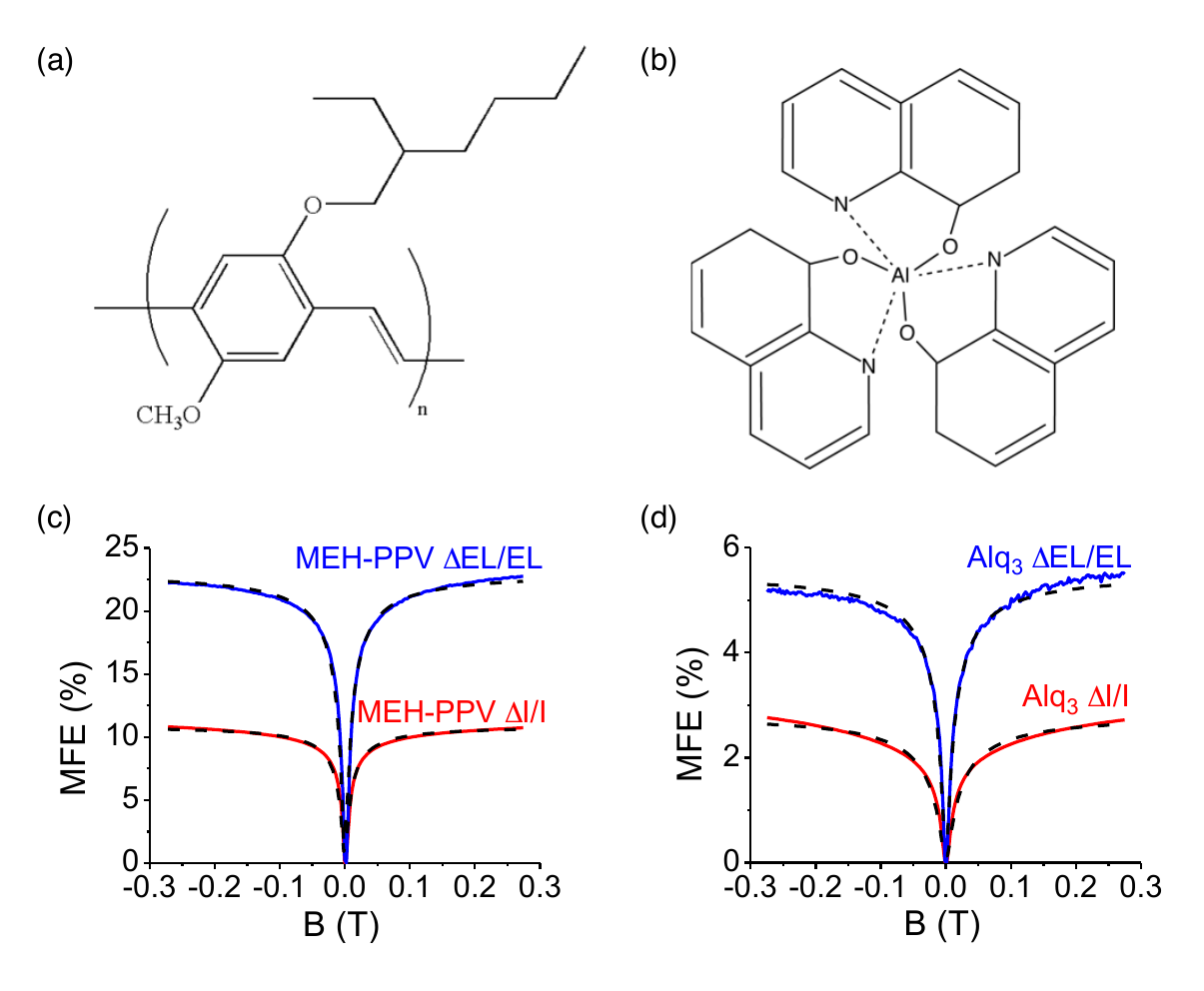}
\caption{\label{fig3aa}
{Molecular structures of  MEH-PPV (a) and Alq$_3$ (b) that have been used in control experiments. Typical magnetconductance (red) and magnetoelectroluminescence (blue) in  MEH-PPV (c) and  Alq$_3$ (d) control devices that are dominated by exciton emission. The dashed black lines represent fits to the non-Lorentzian lineshape (see text)}
}
\end{figure}

The control  devices, with organic luminescent or resistive layers  of Poly[2-methoxy-5-(2-ethylhexyloxy)-1,4-phenylenevinylene] (MEH-PPV, see Fig.~\ref{fig3aa}(a)) or Tris(8-hydroxyquinolinato)aluminium (Alq$_3$, see Fig.~\ref{fig3aa}(b)), consist of a thin film of the organic semiconductor sandwiched between a top and bottom electrode. The indium tin oxide (ITO, 100 nm) coated glass substrates were obtained from Delta Technologies. The substrates were cleaned in an ultrasonic bath using acetone, methanol and isopropanol followed by oxygen plasma cleaning. The conducting polymer poly(3,4-ethylenedioxythiophene)-poly(styrenesulfonate) (PEDOT), purchased from Ossila Ltd., was spin coated at 4000 revolutions per minute (rpm) on top of the ITO to provide an efficient hole injecting electrode. All other manufacturing steps were carried out in a nitrogen glove box. The MEH-PPV (used as purchased from Sigma Aldrich) films were spin-coated from a toluene solution with concentration 5 mg/mL. The Alq$_3$ (used as purchased from Sigma Aldrich) films were thermally evaporated in high vacuum onto the PEDOT covered substrate. The organic semiconductor layer thickness was $\approx$ 150 nm. The cathode layer consisting of calcium and aluminum was deposited by thermal evaporation or e-beam evaporation at a base pressure of $10^{-7}$ mbar on top of the organic semiconductor layer. The active device area was roughly $0.6\times0.6$ mm$^{2}$.

MFE measurements were performed with the device mounted inside a closed-cycle He cryostat placed between two poles of an electromagnet.  All data shown in this article were measured at room temperature, unless specified otherwise. The devices were driven at either a constant voltage V or a constant current I using a Keithley 2400 Source Meter. The particular current and voltage bias levels were chosen in pairs such that they corresponded to similar driving conditions. The electroluminescence intensity was measured using a photomultiplier tube which was shielded from the magnetic field during sweeps. For the optimal device conditioning procedure (see section~\ref{sec:deviceconditioning}), the devices were driven around a current density of 30 mA/cm$^{2}$ for 40 mins. Several MC/MEL traces were taken during the conditioning process to monitor the conditioning effectiveness and to find the optimal conditioning point.

The MFE can be detected either by measuring $I$ as a function of $B$, or, alternatively  by measuring the electroluminescence (EL) as a function of $B$. These two measurements are performed simultaneously  [see Fig.~\ref{fig2ab}(a)] when either the applied device voltage is kept constant (measuring a change $\Delta I(B)$ and $\Delta EL(B)$), or when the device current is kept constant (measuring a change in applied voltage $\Delta V(B)$ and $\Delta EL(B)$ ). The MFE, $\Delta x/x$, in all these quantities is defined as $[x(B)-x(0)]/x(0)$.

Figure~\ref{fig3aa} shows the magnetic field effects in MEH-PPV and Alq$_3$ devices.  These results will serve as a reference point for the discussion of the new types of MFE observed in the TADF devices. MEH-PPV and Alq$_3$ were chosen because their MFEs have been extensively characterized in the OMAR literature\cite{Kalinowski2003,Mermer:2005:PRB} and device conditioning was first reported for MEH-PPV devices\cite{Niedermeier:2008:PRB}.  For typical MFEs the dependence of conductivity (or luminescence) on magnetic field is commonly  either Lorentzian, $\Delta I(B)/I \propto B^2/(B^2+B_0^2)$, or follows a specific non-Lorentzian form, $\Delta I(B)/I \propto B^2/(|B|+B_0)^2$, for the change in the current $I$ where $B_0 \approx 5$ mT\cite{Mermer:2005:PRB}. We find that the data  in Fig.~\ref{fig3aa} can be accurately fitted by the non-Lorentzian expression. 

\section{Magnetic Field Effects in Unconditioned Exciplex Devices}\label{unconditioned}
\begin{figure}
\includegraphics[width=\linewidth]{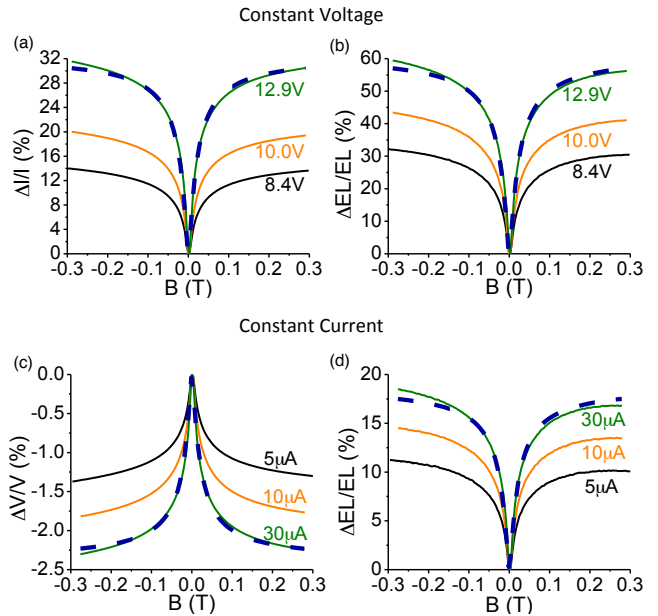}
\caption{\label{fig3a}
\sloppy
\noindent
(a) Magnetoconductance (MC) and (b) magnetoelectroluminescence (MEL) before device conditioning for three representative constant voltages. (c)~Magnetovoltage and (d) magnetoelectroluminescence before device conditioning for three representative constant currents.
The same color is used throughout the figure to designate measurements with a comparable current-density. Thick dashed lines are non-Lorentzian (as defined in text) fits to the largest displayed curve in each panel.}
\end{figure}


Figure~\ref{fig3a} shows measurements of MFE in as-prepared (unconditioned) devices for both constant voltage and constant current measurements. Figure~\ref{fig3a}(a) and~(b) shows that these devices exhibit sizable magnetoconductance and magnetoluminescence effects of up to 30\% and 60\%, respectively, for constant voltage measurements, whereas the effects for constant current measurements are smaller.
The data in Fig.~\ref{fig3a} is well-fit by the non-Lorentzian expression, as Fig.~\ref{fig3aa} was.
The large (60\%) positive MEL is the largest reported in organic semiconductors and remarkably, is completely at odds with the hyperfine mechanism where MEL $\sim$ $-$50\% is predicted for exciplexes (where $k_S > k_T$).

However, our observations are consistent with the $\Delta g$ mechanism operating; the dominance of the $\Delta g$ mechansim is not surprising for MEL and MC in TADF materials because (1) the $g$-factors for two polaron spins ($\bm{S}_1$ and $\bm{S}_2$) on different adjacent molecules ($g_1$ and $g_2$) are expected to vary much more than for identical molecules and (2) McConnell's rule\cite{McConnell1958} states that $B_{0HF} \approx (2-3mT)/\sqrt{N}$ in molecules with $N$ hydrogen atoms. Whereas our molecules in Fig.~\ref{fig1}(a) do not contain many more hydrogens than other molecular organic semiconductors where hyperfine spin-mixing is dominant, the rapid dissociation and association of the exciplex effectively further delocalizes the polarons and reduces the hyperfine interaction.\cite{Dodin2013}

\begin{figure}
\includegraphics[width=\linewidth]{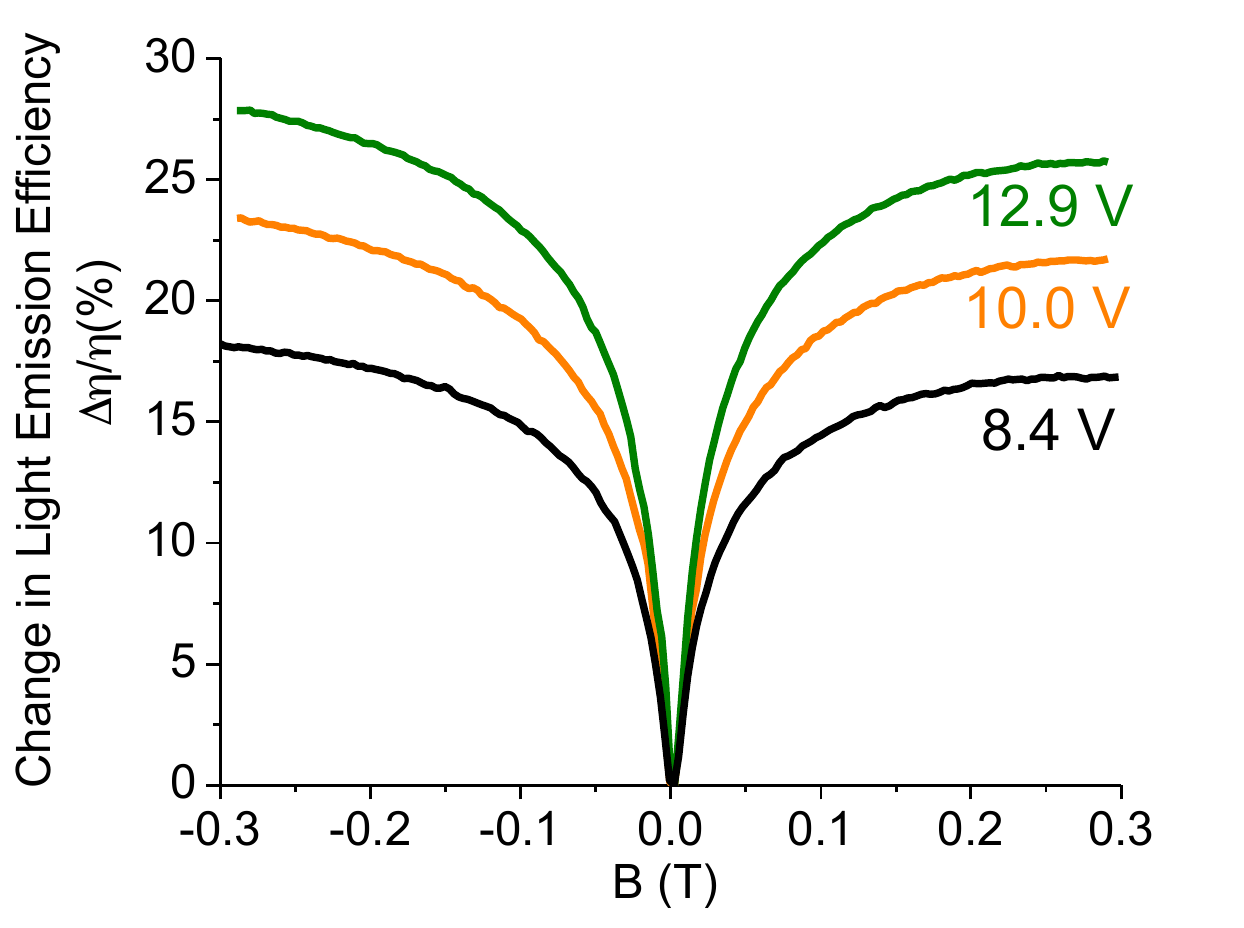}
\caption{\label{figMELMC}
\sloppy
Magnetic field effect on electroluminescence efficiency $\eta$ in the pristine device for three representative constant voltages. The curves shown here are for the same device and experimental data set as in Fig.~\ref{fig3a} (a) and (b).
}
\end{figure}

The data shown in Fig.~\ref{fig3a} are fit to Eq.~(\ref{eq:MELeq}). During this fit $k_S$ and $\Delta g$ are kept constant. $k_S$ has been determined to be $3\times 10^6$~s$^{-1}$ for a similar combination of materials\cite{Goushi2012}. In other organic materials, $\Delta g$  has been found to be on the order of $10^{-3}-10^{-4}$.\cite{Graeff2005, Yu2012} We fix $\Delta g=10^{-4}$, corresponding to the low end of that spectrum.  The crucial fitting parameters are $k_D$ and $G_T/G_S$. $k_D \sim 10^4$~s$^{-1}$, whereas $k_T$ is considerably smaller than $k_D$ and can therefore be assumed zero. Assuming that $G_T=G_S$ at high temperature, the value $G_T/G_S=0.19$ we find from our fit corresponds to an activation energy of $42$~meV, consistent with the TADF behavior.

The maximum change in the electroluminescence with an applied field, relative to the zero-field electroluminescence, measured for a constant voltage (MEL$|_V$) is 0.62, whereas when measured for a constant current (MEL$|_I$) is 0.19. The magnetic-field width of the curves at constant voltage and constant current are very similar, corresponding to 10.8~mT at constant voltage and 10~mT at constant current.

The observation that the MEL effect is considerably larger than the MC effect implies that an applied magnetic field leads to an enhancement in the device's electroluminescence efficiency. The electroluminescence power efficiency, defined as the light output power divided by the electrical input power, is proportional to $EL(B)/I(B)$ at constant voltage, and whereas the EL increases by up to 60\%, only 30\% more current is drawn from the voltage source. Because both quantities, $EL(B)$ and $I(B)$, are measured simultaneously in our experiments, we can readily plot the MFE on  internal electroluminescence quantum efficiency $\eta$ vs. $B$ (see Fig.~\ref{figMELMC}). Our devices are designed for studies of the MFE  and are not high performance OLEDs, therefore a comparison of the external quantum efficiency of our devices to those of highly optimized OLEDs reported in the literature would not be meaningful. The same material combination used here, however, has been utilized in highly efficient OLEDs\cite{Goushi2012}. 

TADF relies on a ``reverse intersystem crossing'' as its underlying mechanism, which in turn is just an umbrella term for spin-dependent singlet-triplet mixing interactions. We have therefore uncovered that the $\Delta g$-mechanism studied here makes a significant (and possibly dominant) contribution to this ``reverse intersystem crossing". Two interesting future research directions therefore become apparent: (i) the use of external $B$-fields to enhance the efficiency of TADF and ultimately exciplex OLEDs and (ii) the use of MFE spectroscopy to study the nature of the spin-dependent processes that are responsible for the ``reverse intersystem crossing''.

\section{MFE Enhancement by Device Conditioning} \label{sec:deviceconditioning}

\begin{figure}
\includegraphics[width=\linewidth]{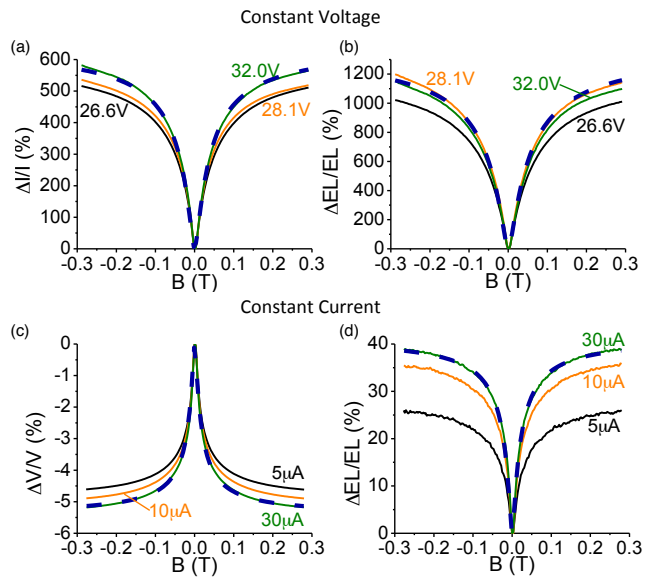}
\caption{\label{fig3b}
\sloppy
\noindent
(a) Magnetoconductance and (b) magnetoelectroluminescence after device conditioning for three representative constant voltages. (c)~Magnetovoltage and (d) magnetoelectroluminescence after device conditioning for three representative constant currents.
The same color is used throughout the figure to designate measurements with a comparable current-density. Thick dashed lines are non-Lorentzian (as defined in text) fits to the largest displayed curve in each panel.
}
\end{figure}

The effects seen in Fig.~\ref{fig3a} are amongst the largest reported in the OMAR literature so far, but Fig.~\ref{fig3b}{(a) and~(b)} show that the effects, measured in the same device, increase spectacularly to over 500\% MC and 1000\% MEL, after so-called ``device conditioning''  was performed. The effect of electrical conditioning of (excitonic) OMAR devices was  reported by Niedermeier et al.\cite{Niedermeier:2008:PRB}, who found that the MC can be increased from $\approx$ 1\% to $\approx$ 15\%~{in MEH-PPV devices}. Our conditioning procedure is similar to theirs, and consists of operating the device over a period of time at a relatively high current density. We note for later discussion that the increase in the constant voltage MFE is much larger than the increase in the constant current MFE, although even the constant-current MFE increase is over a factor of 2.

\begin{figure}
\includegraphics[width=\linewidth]{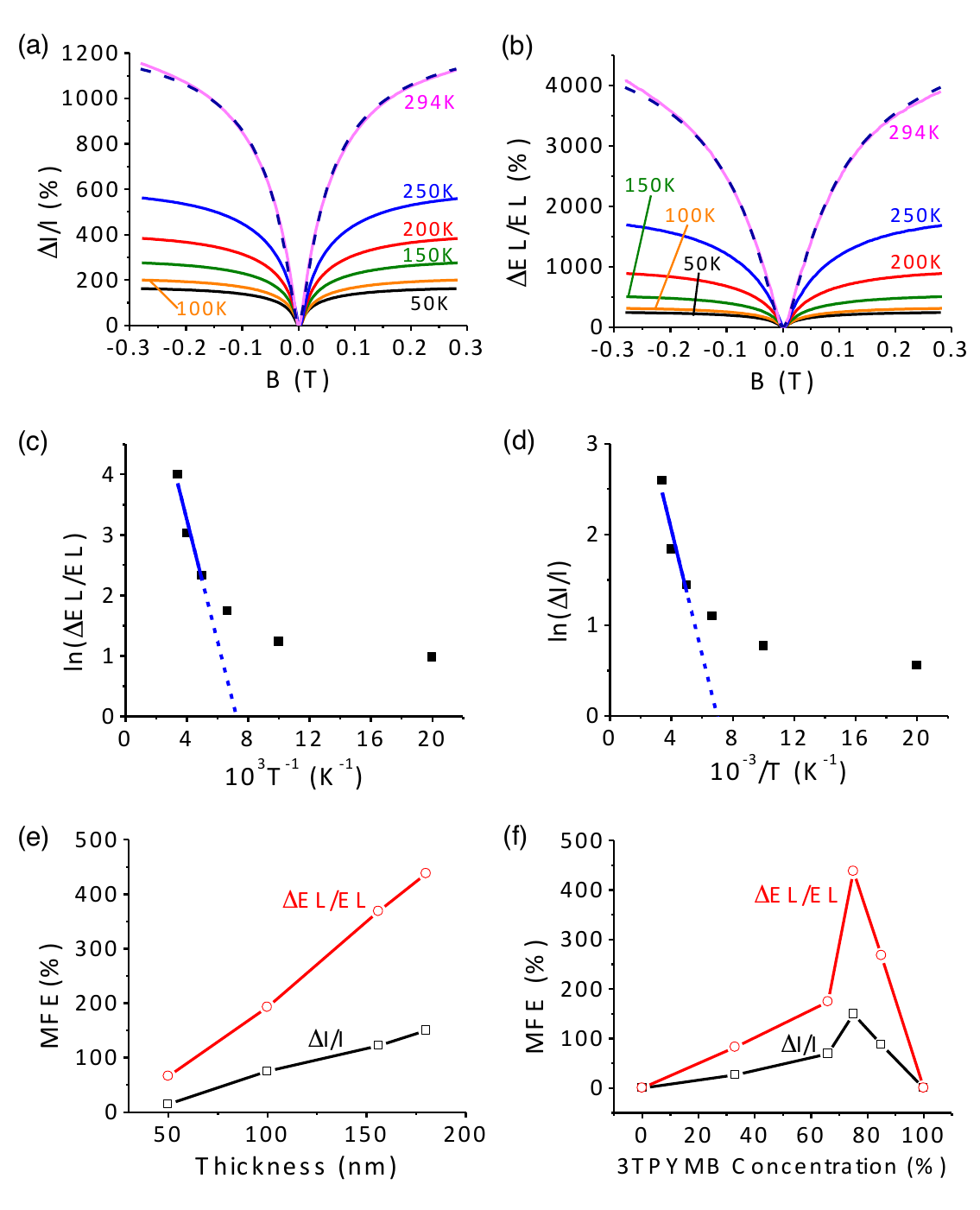}
\caption{\label{fig4a}
\sloppy
\noindent
\rm{({a}) and ({b}) Temperature dependence of magnetic-field-effects in optimally conditioned device. Dashed lines are non-Lorentzian (as defined in text) fits. ({c}) and (d) Arrhenius plot of the saturation magnetic field effect extracted from the data in (a) and ({b}), respectively. Solid lines are fits to an activated Boltzmann law indicating the range of data points used in the fit, whereas the dashed lines are extrapolations to lower temperatures. ({e}) Dependence of the saturation magnetoconductance ($\Delta I/I$) and magnetoluminescence ($\Delta EL/EL$) versus the thickness of the co-evaporated layer, whereas ({f}) plots the same quantities versus the composition of the co-evaporated layer.}
}
\end{figure}

We now present a detailed description of the measurements made in the conditioned device that exhibits the largest MFE. These values were reproduced in several devices.
Fig.~\ref{fig4a} shows MFE results as a function of various additional experimental parameters and device fabrication specifications, including temperature (Fig.~\ref{fig4a}(a)-(d)), thickness (Fig.~\ref{fig4a}(e)) and composition (Fig.~\ref{fig4a}(f)). The devices shown in this figure follow the same design as the device used in the previous figure but were fabricated from a different batch of materials {(see  section~\ref{sec:devicefabrication})}. The effects we report are not due to  local shorts or pinholes in the device, as even the optimally conditioned device reported here is stable and operates at a noise level equal or better to that of the unconditioned device (see videos in Supplementary Information\cite{SupMat}). Fig.~\ref{fig4a}(a) and~(b) shows that an even more dramatic increase in MC and MEL upon conditioning has been achieved in this device. $B_0$ increases to up to $\approx 50$ mT upon conditioning. This is much larger than the expected hyperfine strength, $B_{0hf}$, in our materials (also compare with Fig.~\ref{fig3aa}).


Only those triplet exciplexes activated to near the singlet level (dashed lines, $T^{\ast}$, in Fig.~\ref{fig1}(c)) can be involved in the conversion to singlets. Hence, $G_T < G_S$ in general except at the highest temperatures when the two are equal. From this argument we expect the MFEs to increase with a rise in temperature (due to the increase in $G_T$), in agreement with experiment (Fig.~\ref{fig4a}(a) and~(b)). At very low temperatures the experimental data does not follow a simple Boltzmann law, but at temperatures above 200 K it agrees with a Boltzmann dependence with an activation energy $\Delta_{ST} \approx$ 60 - 80 meV (Fig.~\ref{fig4a}(c) and~(d)). This compares favorably with literature values\cite{Goushi2012} obtained from spectroscopic studies covering a similar temperature range as our Boltzmann fit.
Fig.~\ref{fig4a}(g) shows that the magnitude of the effect increases with increasing device thickness, and Fig.~\ref{fig4a}(h) shows that the effect is maximal for $x=0.75$ in the mixed layer. Adachi~{\it et al.} found that $x=0.75$ leads to devices with maximum TADF efficiency.\cite{adachi2012delayed}

\subsection{Relationship between constant-current MC and constant-voltage MC}
\label{section2}

 Fig.~\ref{fig1}(e) shows that the maximum predicted constant-current MEL is 100\% which is consistent with our constant current measurements, which do not surpass 40\% [see Fig.~\ref{fig3b}(d)]. Experimentally, the largest MFEs are observed in the constant-voltage mode, where the effects easily surpass 100\%. This is possible because the material combination in conjunction with the specified conditioning procedure yields highly nonlinear I-V and EL-V curves.

Converting between the constant voltage and current modes of MC ($\Delta I/I$ and $\Delta V/V$, respectively) can be carried out once the I-V relationship is determined. We write the I-V characteristics for magnetic field off and on as
\begin{equation}
I_0 = V^{\alpha},\qquad I_B = (V-\Delta V )^{\alpha}
\end{equation}
and by the definition of MC, we write:
\begin{equation} \label{equ:constantItoVconversion}
MC = \frac{I_B - I_0}{I_0} =  \Big(1-\frac{\Delta V}{V}\Big)^{\alpha}-1
\end{equation}
which is an exact expression.
By the binomial series expansion, we obtain leading order terms
\begin{equation}
MC \approx -\alpha \frac{\Delta V}{V} +\frac{\alpha}{2}(\alpha - 1) \left (\frac{\Delta V}{V} \right )^2 - ...
\end{equation}

The exponent of the nonlinearity, $\alpha$, has been associated with trap-limited transport when the trap energies follow an exponential density of states.\cite{Mark1962} The reader should note that the MC at constant voltage increases linearly (to first order) with $\alpha$. Therefore, any process that increases $\alpha$ will likely increase the observed MC. This ``$\alpha$-enhancement" allows MFE at constant voltage to significantly exceed the 100\% maximum predicted by our theory for the constant-current effect.

\subsection{Effect of electrical conditioning on $IV$ and $EL-V$ characteristics}

\begin{figure}
\includegraphics[width=\linewidth]{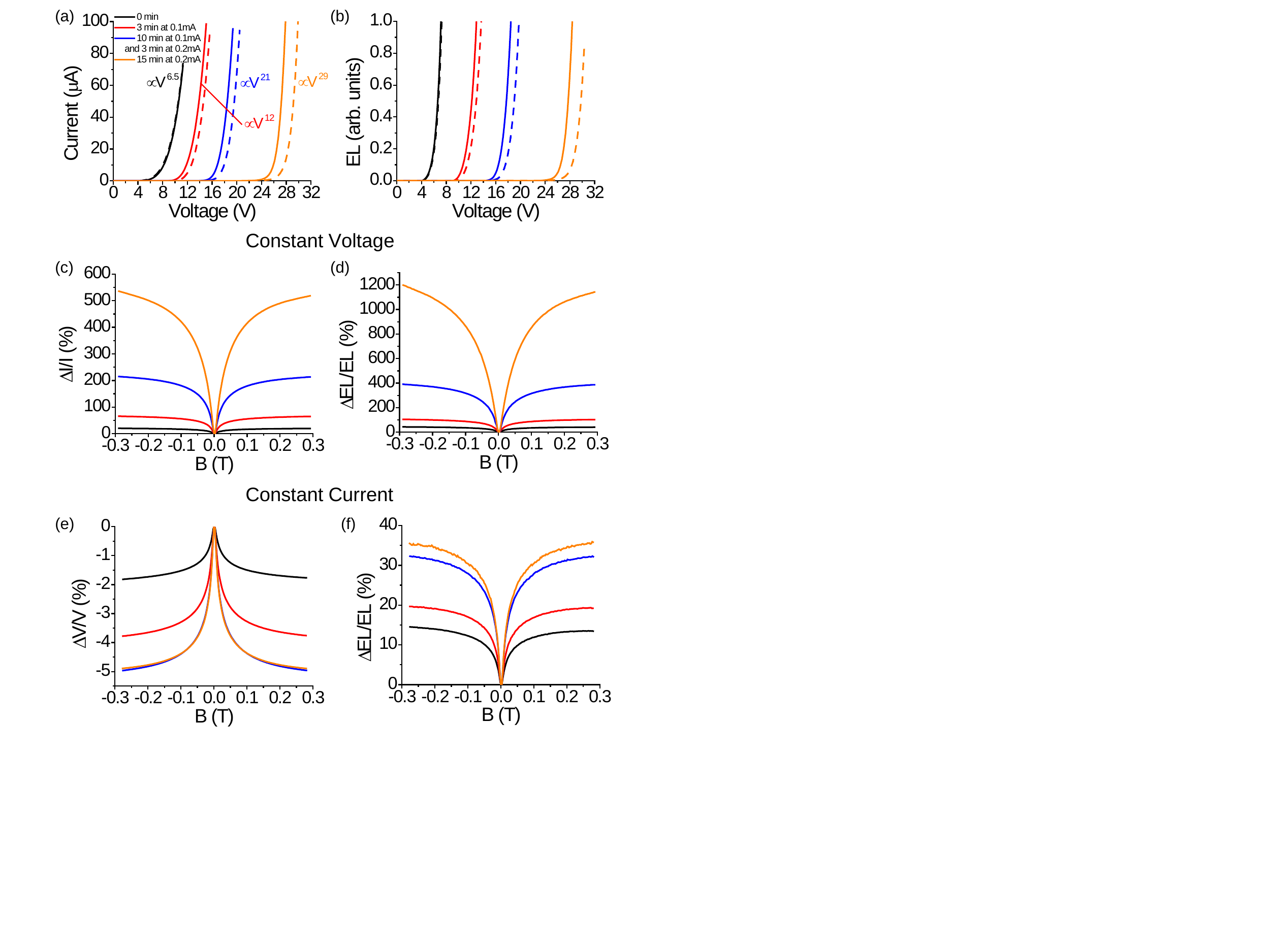}
\caption{\label{fig5}
Current-voltage ({a}) and electroluminescence-voltage ({b}) curves for different stages of device conditioning (protocols of the conditioning are specified as labels to the curves of  ({a}); other panels use same color coding).  Dashed (solid) lines are measurements at $B = 0$ ($0.3$) T.
Constant voltage ({c}), ({d}) and constant current ({e}), ({f}) magnetic-field-effect measurements for varying device conditioning.}
\end{figure}

We now examine the effect of electrical conditioning on the current-voltage (I-V) and electro-luminescence-voltage (EL-V) characteristics of the device.
Figure~\ref{fig5} shows that $\alpha$ increases with device conditioning. Figure~\ref{fig5}(a) shows the I-V characteristics of the pristine device (``0 min") and the same device after several device conditioning operations (the current levels and conditioning times are assigned). Each conditioning step is performed on the device that was already conditioned by the cumulative prior steps. Data curves measured with an applied B-field of 0.3 T are shown as solid lines, and the zero-field characteristics are shown as dashed lines. The figure shows that the I-V curves shift to higher voltages, with an onset voltage of about 5 V and 25 V for the pristine and maximally conditioned device, respectively (further device conditioning would lead to a rapid device degradation and ultimately device failure). Both the I-V and EL-V traces are approximately given by a power law with exponent $\alpha$.  At the same time it is observed that the dependence of $I$ on $V$ becomes increasingly non-linear ($I \propto V^\alpha$) with the exponent $\alpha$ increasing from approximately 6.5 to 30. Comparing the solid and dashed lines, we find that phenomenologically the effect of the applied magnetic field is a parallel-shift of the I-V traces by a negative amount $\Delta V$. Similar behavior characterizes the EL-V traces shown in Fig.~\ref{fig5}(b).

\begin{table*}[]
\centering
\caption{Effect of various device conditioning procedures on a square device $0.6$~mm on a side (see Section~\ref{sec:devicefabrication} for details). The row number of the table signifies increasing conditioning times and /or currents. Specifically ``Before Device Conditioning" is for zero conditioning time, ``First Device Conditioning" is for 3 mins conditioning at 0.1~mA, ``Second Device Conditioning" refers to an additional conditioning of 10 mins  at 0.1~mA plus an additional 3~mins at 0.2~mA, whereas ``After Device Conditioning" signifies an additional conditioning for 15 mins  at 0.2~mA resulting in the optimally conditioned device. The first four columns contain directly measured quantities: MEL$|_{V}$ (MEL$|_{I}$) is the maximum MEL measured for a constant voltage (current), $B_0|_V$ ($B_0|_I$) is the half-width at quarter-maximum and is also the parameter entering into the non-Lorentzian empirical law. The next two columns show quantities that can be calculated from the measured data without reference to any theory: $\alpha$ is the exponent of the device's non-linear I-V characteristics (see Section~\ref{section2}) and $\eta$ is the relative electroluminescence efficiency ($\eta \equiv 1$ for the pristine device). The remaining quantities were extracted from fits to our theory (see text).}
\label{my-table}
\begin{tabular}{ c   c   c   c   c   c   c   c   c   c   c }
\hline\hline
\begin{tabular}[c]{@{}c@{}}Device Conditioning\end{tabular} & MEL$|_{V}$ & MEL$|_{I}$ & $B_0$ $|_ {V}$ (mT) & $B_0$ $|_ {I}$ (mT) & $\frac{MEL|_{V}}{MEL|_{I}} $ & $\alpha$ & $\eta$ & $k_D$ (s$^{-1}$) & $\frac{G_T}{G_S}$ & $\Delta_{ST}$ (meV) \\ \hline
Before Device Conditioning                                                     & 0.62  & 0.19  & 10.8     & 10       & 3.27    & 13.4  & 1     & $1\times10^{4}$ & 0.19  & 42  \\ 
First Device Conditioning                                                        & 1.15  & 0.23  & 11.3     & 9.5      & 5.00    & 16.2  & 0.38  & $9\times10^{3}$  & 0.23  & 37  \\ 
Second Device Conditioning                                                        & 4.00  & 0.35  & 14.5     & 9.5      & 11.4    & 24.1  & 0.16  & $9.2\times10^{3}$    & 0.35  & 26  \\ 
After  Device Conditioning                                                     & 13.19 & 0.41  & 26.7     & 9.8      & 32.2    & 32.0  & 0.04  & $9.6\times10^{3}$   & 0.42  & 22  \\ \hline\hline 
\end{tabular}
\end{table*}

\subsection{Effects of device conditioning on magnetic field effects}

Figure~\ref{fig5}(c) and (d) show the effect of device conditioning on MC and MEL measured at a constant applied voltage at several different stages of device conditioning (using the same color coding as for Fig.~\ref{fig5}(a)), and Fig.~\ref{fig5}(e) and~(f) show the corresponding data for constant current measurements. Figure~\ref{fig5}(c) and~(d) show that the MC magnitude dramatically increases from about 20\% for the pristine device, to about 500\% for the maximally conditioned device, and the EL magnitude increases from about 50\% to about 1200\%. Device conditioning has a significantly less dramatic effect on the data shown in Fig.~\ref{fig5}(e) and~(f), which are for measurements at a constant current.

{Table~\ref{my-table} shows the fitting results for different amounts of device conditioning. We fit the constant current measurements with Eq.~\ref{eq:MELeq} by holding $k_S$ and $\Delta g$ constant, to the same values used in Sec.~\ref{unconditioned} for the unconditioned devices, and setting $k_T=0$. We fit all data sets to the non-Lorentzian expression. We find that the non-Lorentzian lineshape results in excellent fits to the measured data, and the resulting fitting parameters are listed in {Table~\ref{my-table}}. 

Noteworthy aspects of the fitting results are: (i) $\alpha$ increases significantly from 13 to 32 upon conditioning, and a corresponding increase of the MEL$|_V$ measured at constant voltage over the MEL$|_I$ measured at constant current is observed. (ii) The electroluminescence efficiency, $\eta$ decreases to 4\% of the pristine value upon device conditioning. (iii) the curve width for constant voltage, $B_0|V$, increases upon conditioning but the curve width for constant current, $B_0|I$ does not. (iv) The exciplex dissociation rate $k_D$ appears to be largely insensitive to conditioning, whereas conditioning moves the triplet-to-singlet branching ratio to favor triplet formation more. In terms of our theory, this  indicates a reduction of the activation energy $\Delta_{ST}$. Because $\Delta_{ST}$ is a sensitive function of the electron-hole wavefunction overlap, a change in this quantity is to be expected if device conditioning results in changes to film morphology, molecular packing or electron-hole localization along the current flow paths. The increase in resistance associated with conditioning implies that conducting sites or electron-hole recombination sites are farther apart, which also is consistent with the reduction in $\Delta_{ST}$. We note that further increase by conditioning in the MEL at constant current (which is connected directly to the $\Delta_{ST}$ in our theory and Eq.~(\ref{eq:MELeq})) should be possible since the regime of $\Delta_{ST}\ll k_BT$ has not yet been reached. The increase of $B_0|V$ while $B_0|I$ is unchanged in conditioning appears related to the increase of $\alpha$ with conditioning.

Although the reduction of the electroluminescence efficiency is not a problem for magnetosensor applications, it is undesirable if the device is to be used as a magnetosensitive light-emitting device. However, there is no known reason why this reduction in efficiency should be a necessary companion to large MFE. Procedures for achieving larger MEL without reducing $\eta$ may be found in the future, once the microscopic mechanism of conditioning is better understood. For example, we speculate that conditioning decreases $\Delta_{ST}$ by increasing the average separation of the electron donor molecule and the electron acceptor molecule, which simultaneously increases the resistance. Finding molecule pairs with smaller $\Delta_{ST}$ but higher conductivity could lead to larger MELs without compromising $\eta$. The exciplex model here will produce the largest MFEs  for a system with a $\Delta_{ST}\ll k_BT$, which is not a regime we reach even with conditioning. 

\begin{figure}
\includegraphics[width=\linewidth]{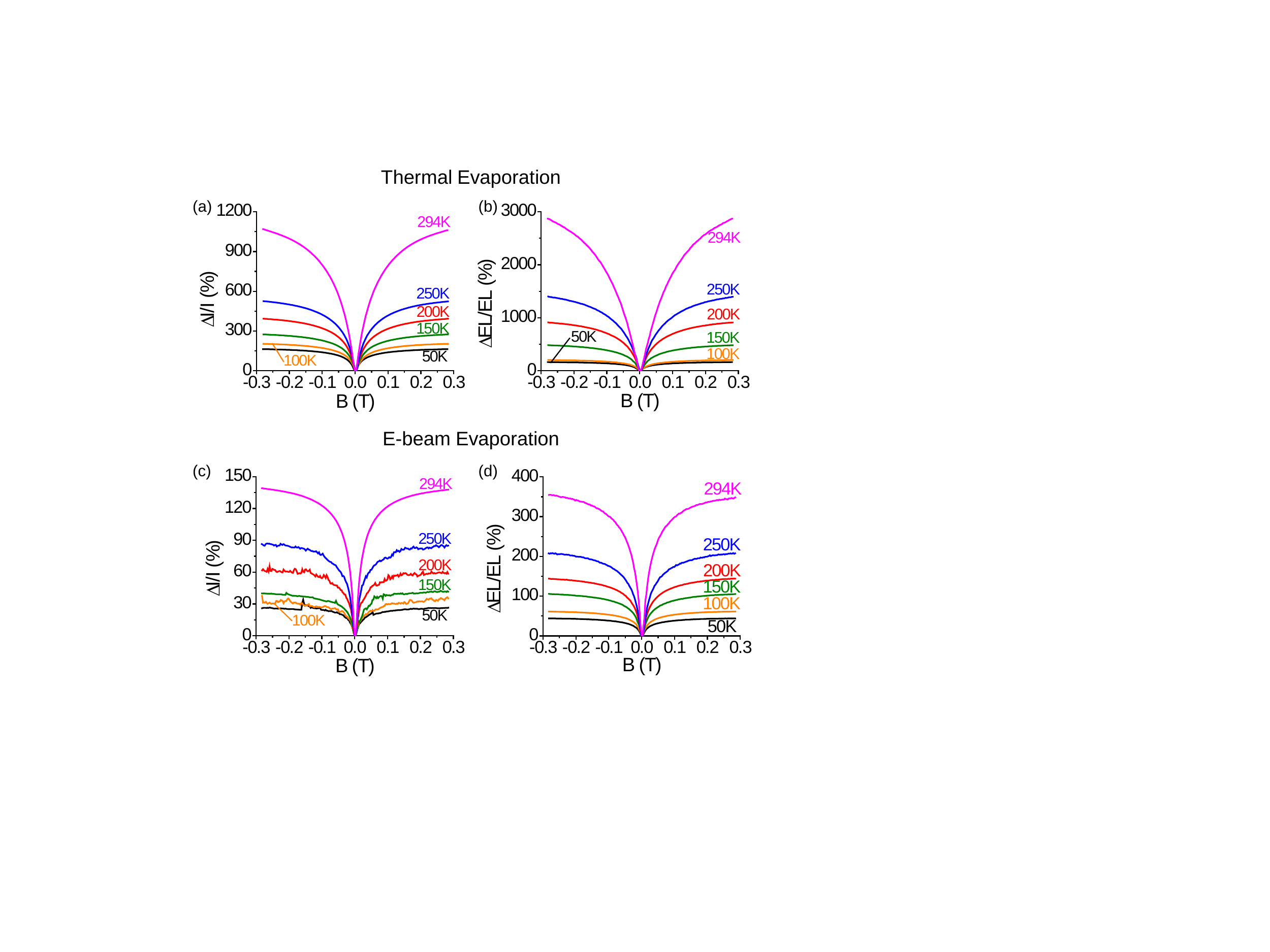}
\caption{\label{fig6}
{Magnetoconductance and magnetoelectroluminescence in TADF exciplex based devices that differ in the fabrication procedure used to deposit the top metal electrode.} \rm{(a), (b) Devices whose top electrode was fabricated by thermal evaporation. These devices have been maximally conditioned. (c), (d) Devices whose top electrode was fabricated by e-beam evaporation. These devices are not sensitive to device conditioning.}
}
\end{figure}

It was shown previously\cite{Rybicki:2012:PRL} that the exposure of the organic layer to X-ray bremsstrahlung that is generated during the e-beam evaporation process can significantly increase the MFE. It is therefore interesting to investigate whether devices fabricated using e-beam evaporation for the top electrode/cathode layer instead of thermal evaporation show an MFE enhancement similar to device conditioning. Figure~\ref{fig6}(a) and~(b) show MC and MEL measurements for the maximally conditioned thermally fabricated device as a function of temperature. We now compare these data for the thermally conditioned device to analogous data in an e-beam fabricated device (Fig.~\ref{fig6}(c) and~(d)). Whereas the MFE in e-beam fabricated devices is large compared to unconditioned devices fabricated by thermal evaporation (see Sec.~\ref{unconditioned}), it is a much smaller MFE than the thermally fabricated device after device conditioning (see Sec.~\ref{sec:deviceconditioning}). In fact we found that the e-beam devices are no longer sensitive to the device conditioning process. For this reason, e-beam fabricated devices are the most suitable for studies of the dependence of MFE on various parameters (such as layer thickness, layer composition, etc.) where an additional dependence on current density is not desired, as it would complicate the analysis. We {therefore} used e-beam devices {for Fig.~\ref{fig4a}(e),(f)}.

As a final comparison we show the stability of the conditioning procedure. Figure~\ref{fig4b} shows that the device does not return to its unconditioned value after resting for 12~hours. Further studies of the evolution of the magnetic field response with rest time will be the subject of future work.

\begin{figure}
\includegraphics[width=\linewidth]{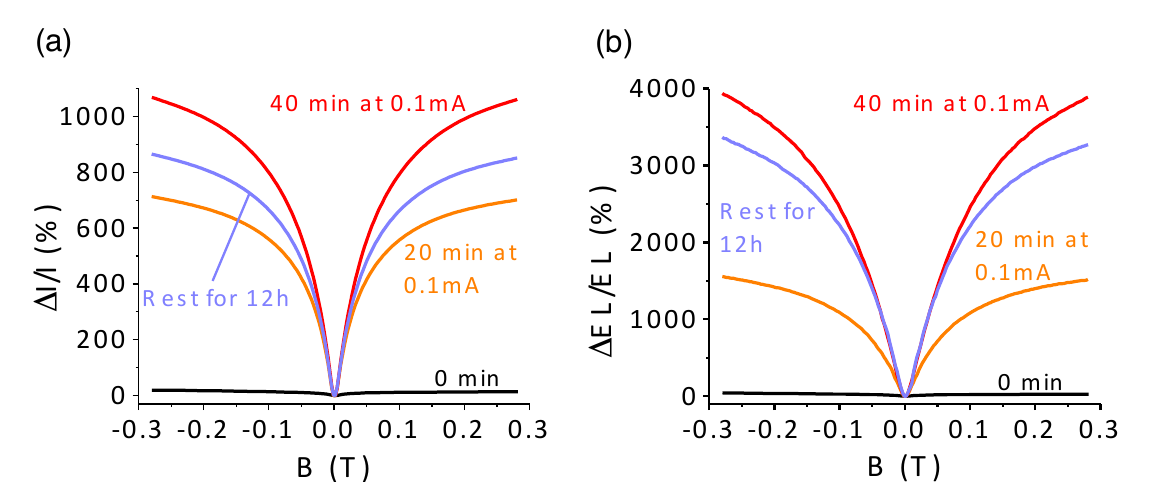}
\caption{\label{fig4b}
\sloppy
\noindent
(a) and (b) Conditioning dependence of the device studied in Fig.~\ref{fig3b}. The blue line is for the optimally conditioned device after a rest period of 12h and shows that the effect does not return to the pristine value.
}
\end{figure}

\section{Comparison of magnetic field effects in exciplex devices to excitonic devices}
\label{section3}

\label{section4}
\begin{table}
\small{
\caption{\label{S2}
\rm{Experimental values for $\Delta V/V$ and MC at different currents in MEH-PPV, Alq$_3$  and m-MTDATA:3TPYMB devices. $\alpha$ is obtained from fitting the IV curves and the expected MC is calculated from Eq.~\ref{equ:constantItoVconversion} in Sec.~\ref{section2}.}}
\rm{
\begin{center}
\begin {tabular} { c c c c c }
 \hline
 \hline
 Description & $\alpha$ & $\Delta V/V$ & expected MC & MC\\ \hline
 \multicolumn{5} { c }{MEH-PPV before  Device Conditioning}\\ \hline
 5 $\mu$A & 7.5 & -0.0088 & 0.068 & 0.074\\
 10 $\mu$A & 7.5 & -0.0085 & 0.065 & 0.07\\
 20 $\mu$A & 7.5 & -0.0081 & 0.063 & 0.068\\ \hline
 \multicolumn{5}{ c }{MEH-PPV after  Device Conditioning}\\ \hline
 1 $\mu$A & 8.1 & -0.0166 & 0.142 & 0.131\\
 10 $\mu$A & 8.1 & -0.0143 & 0.122 & 0.109\\
 100 $\mu$A & 8.1 & -0.0085 & 0.071 & 0.071\\ \hline
 \multicolumn{5}{ c }{Alq$_3$ }\\ \hline
 1 $\mu$A & 7.2 & -0.0056 & 0.041 & 0.041\\
 10 $\mu$A & 7.2 & -0.0041 & 0.03 & 0.028\\
 100 $\mu$A & 7.2 & -0.0033 & 0.024 & 0.024\\ \hline
 \multicolumn{5}{ c }{m-MTDATA:3TPYMB at 30 $\mu$A}\\ \hline
 Before Device Conditioning & 13.43 & -0.024 & 0.372 & 0.325\\
First Device Conditioning & 16.24 & -0.039 & 0.857 & 0.695\\
 Second Device Conditioning & 24.12 & -0.052 & 2.425 & 2.094\\
After Device Conditioning & 32 & -0.054 & 4.418 & 6.436\\
 \hline
 \hline
\end{tabular}
\end{center}
}
}
\end{table}

\begin{figure}
\includegraphics[width=0.98\linewidth]{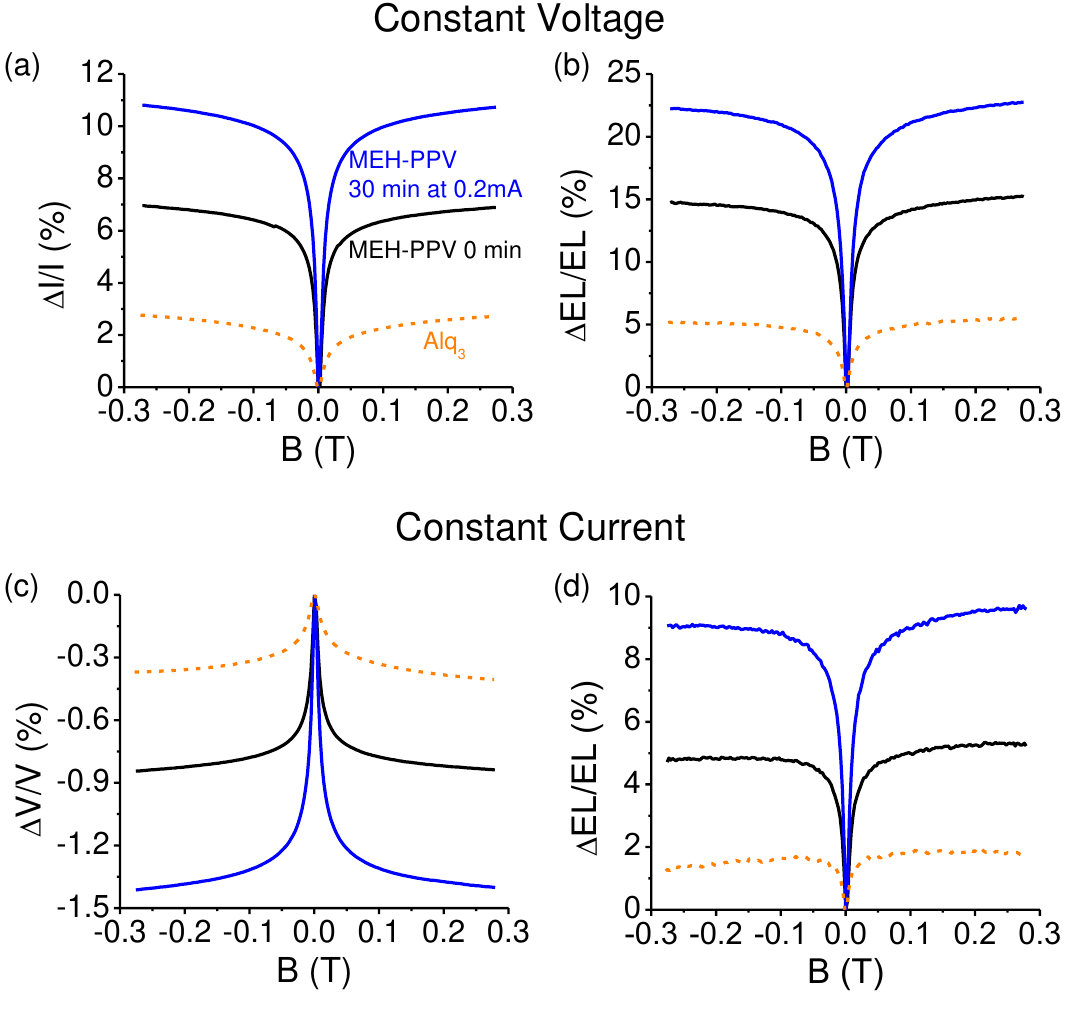}
\caption{\label{fig7cf}
Magneto-transport and magneto-luminescence in MEH-PPV and Alq$_3$  devices. Magneto-transport and magneto-luminescence at constant voltage (a), (b) and constant current (c), (d) before and after device conditioning in MEH-PPV and Alq$_3$  devices. }
\end{figure}

In this section we compare MFE in our exciplex system to the (previously known) excitonic MFE. The simple, single layer excitonic devices studied  are not optimized to be efficient OLEDs, but serve as a reference for our MFE studies. 
 Figure~\ref{fig7cf}(a)-(d) shows that the magnitude of MFE in MEH-PPV also increases with electrical conditioning, but that the MFE before and after conditioning are both much smaller than in the exciplex devices.
The MFE of the Alq$_3$ device is not sensitive to the electrical device conditioning, but it has previously been shown that a kind of device conditioning can be achieved by x-ray exposure\cite{Rybicki:2012:PRL}. The value of $\Delta V/V$ and $\Delta EL/EL$ in MEH-PPV and Alq$_3$ devices at constant current is less than one third compared to the TADF exciplex devices. However, the value of $\Delta I/I$ and $\Delta EL/EL$ at constant votage is smaller than 2\% compared to TADF devices. Moreover, the half-width-at-quarter-maximum of the traces, $B_0$, in MEH-PPV and Alq$_3$ devices is less than 5 mT, but for TADF exciplex devices, $B_0$ can be nearly 50 mT. We note that the reported data are first measured at a constant current from which we obtain $\Delta V/V$ and $\Delta EL/EL$ at constant current. Then we choose a voltage for the constant voltage measurement that results in a similar current flow as for the constant current measurements. In this way, we can perform an accurate, quantitative comparison between constant current and constant voltage measurements. See Table~\ref{S2} for a summary of our results, where the observed value for the exponent $\alpha$, the measured magnetovoltage, the expected magnetoconductance due to the ``$\alpha$-enhancement", and finally the actually measured magnetoconductance are shown. The table shows good agreement between expected and actual magnetoconductance, validating our picture.

\section{Conclusion}

We have shown that TADF-based organic diodes can exhibit immense sensitivity to magnetic fields. 
The measured MFEs are the highest among systems with non-magnetic components. Previous MFEs in organic semiconductors and colloidal quantum dots fail to surpass 20\%\cite{Mermer:2005:PRB, Guyot-Sionnest2007}. Similar MFE values are found in chemical reaction yields\cite{Timmel1998}. However in hybrid perovskites, where $\Delta g$ is responsible for the MFEs, the MFEs are very small, $< 0.5$\%.\cite{Zhang:2015:NatPhys} 
Spin dynamics and the associated MFEs also have  been recently studied for defects between either two non-magnetic leads or one non-magnetic lead and one magnetic lead. In both cases, the MFEs are less than 0.1\%\ and appear to be due to hyperfine interactions\cite{Swartz2014, Inoue2015}. In contrast to systems with magnetic components, in which the magnetic easy axes cause sensitivity to the vector character of the external magnetic field, 
the MFEs of these nonmagnetic TADF-based materials can be expected to depend predominately on the scalar magnitude of the field.
Even for devices with magnetic components --- for instance those exhibiting tunneling magnetoresistance with maximum room temperature observed values $> 600$\% \cite{Ikeda2008} --- our MFEs compare favorably despite other drawbacks inherent in the electronic properties of organic semiconductors when compared to metallic systems. 

Our results imply that magnetic field effects are a sensitive tool for investigating the spin-dependent exciplex physics that lies at the heart of TADF, and also shows that external and internal magnetic fields may serve as a booster of the electroluminescence efficiency in TADF devices.
Our simple conditioning approach proves that immense MFE are achievable in organic semiconductors, even though the microscopic processes that occur upon device conditioning cannot be identified with certainty. We speculate that conditioning introduces chemical changes to the molecules and/or changes to the film's nanoscale morphology that move the electron donors and acceptors effectively farther apart, both decreasing $\Delta_{ST}$ and increasing the resistance. It remains  a challenge to the organic semiconductor field to develop the necessary nanoscale tools to achieve smaller $\Delta_{ST}$'s while retaining higher conductivity, through morphological control during deposition or improved molecular design and targeted chemical synthesis.

\begin{acknowledgments}
We acknowledge advice from J. Shinar on the coevaporation process. This work was supported in part by The University of Iowa Office of the Vice President for Research and Economic Development through the GAP and IFI Funding Programs. This material is based in part on work supported by the U.S. Department of Energy, Office of Science, Office of Basic Energy Sciences, under Award Number DE-SC0014336.
\end{acknowledgments}

\end{document}